\documentclass[onecolumn,superscriptaddress,floatfix,preprintnumbers, nofootinbib,hyperref]{revtex4} 
\pdfoutput=1
\usepackage[colorlinks=true,breaklinks=true]{hyperref}
\usepackage[utf8]{inputenc}
\hypersetup{allcolors=[rgb]{0.0 0.0 0.6},linkcolor=[rgb]{0.75 0.05 0.05}}
\usepackage{amsmath,amssymb}
\usepackage{epsfig}  
\usepackage{graphicx}   
\usepackage{slashed}             
\usepackage{url}
\usepackage{color}
\usepackage{multirow}
\usepackage{placeins}
\usepackage[dvipsnames]{xcolor}
\usepackage{letltxmacro}
\LetLtxMacro{\oldcite}{\cite}
\renewcommand{\cite}[1]{\mbox{\oldcite{#1}}}

\DeclareMathOperator{\GeV}{GeV}

\DeclareMathOperator{\eV}{eV}

\DeclareMathOperator{\kg}{kg}

\DeclareMathOperator{\s}{s}

\DeclareMathOperator{\km}{km}
\DeclareMathOperator{\cm}{cm}

\DeclareMathOperator{\Mpc}{Mpc}
\DeclareMathOperator{\arctg}{arctg}

\newcommand{\bk}{{\bf k}}
\newcommand{\bE}{{\bf E}}
\newcommand{\bB}{{\bf B}}
\newcommand{\bA}{{\bf A}}

\newcommand{\bx}{{\bf x}}

\newcommand{\beq}{\begin{equation}}
\newcommand{\eeq}{\end{equation}}
\newcommand{\bsp}{\begin{split}}
\newcommand{\esp}{\end{split}}

\allowdisplaybreaks

\bibpunct{[}{]}{,}{n}{}{,}

\begin{document}

\title{Dynamical evolution of  axion condensates \\  under stimulated decays into photons
 }

\author{Pierluca Carenza}
\email{pierluca.carenza@ba.infn.it }
\author{Alessandro Mirizzi}
\email{alessandro.mirizzi@ba.infn.it }
\affiliation{Dipartimento Interateneo di Fisica ``Michelangelo Merlin'', Via Amendola 173, 70126 Bari, Italy.}
\affiliation{Istituto Nazionale di Fisica Nucleare - Sezione di Bari,
Via Amendola 173, 70126 Bari, Italy.}
\author{G\"unter Sigl}
\email{guenter.sigl@desy.de}  
\affiliation{II. Institute for Theoretical Physics, Hamburg University
Luruper Chaussee 149, D-22761 Hamburg, Germany}

%=============================================================================

\begin{abstract}
Dark matter axion condensates may experience stimulated decays into photon pairs. 
This effect has been often interpreted as a parametric resonance of photons from the axion-photon coupling, leading to an exponential growth of the photon occupation number in a narrow instability band.
Most of the previous literature does not consider the possible evolution of the axion field due to  the photon growth. 
 We revisit this effect presenting a mean field solution of the axion-photon kinetic equations, in terms of number of photons
 and pair correlations. We study the limit of no axion depletion, recovering the known instability. Moreover, we extend the results including
 a possible depletion of the axion field. In this case we find that the axion condensate exhibits the behaviour of an inverted pendulum.
 We discuss the relevance of these effects for two different cases: an homogeneous axion field at recombination and 
 a localized axion clump and discuss constraints that could result from the induced photon background.
  \end{abstract}

\maketitle

%%%%%%%%%%%%%%%%%%%
\section{Introduction}
\label{sec:1}
%%%%%%%%%%%%%%%%%%%%%%%%%%%%

Axions emerge in relation to the {\em the strong CP problem} of the strong interactions. Indeed, the most elegant solution to this puzzle
is based  on the Peccei-Quinn (PQ) mechanism~\cite{Peccei:1977hh,Peccei:1977ur,Weinberg:1977ma,Wilczek:1977pj}, in which the Standard Model is enlarged with an additional global U(1)$_A$ symmetry, known as the PQ symmetry. The spontaneous breaking of this symmetry leads as associated Nambu-Goldstone boson, 
the axion: 
a low-mass pseudoscalar particle with properties similar to those of neutral pions.
The axion mass is given by~\cite{diCortona:2015ldu,Borsanyi:2016ksw}
%...............................................
\begin{equation}
m_a = \frac{5.7 \,\ \textrm{eV}}{F_a/10^6 \,\ \textrm{GeV}} \,\ ,
\label{eq:ma}
\end{equation}
%.................................................
where {$F_a$} is the axion decay constant or PQ scale.
The axion interactions with photons, electrons, and hadrons are also controlled by the PQ constant  and scale as  {$F_a^{-1}$}.
Notably,
the two-photons interaction of axions 
provides the basis for many possible astrophysical implications and experimental search strategies for these elusive particles. 
It is  described by the  Lagrangian~\cite{Sikivie:1983ip,Raffelt:1987im,Anselm:1987vj}
\beq
\mathcal{L}=-\frac{g_{a\gamma}}{4}\phi\tilde{F}^{\mu\nu}F_{\mu\nu}\ =  g_{a\gamma} \phi \bE\cdot\bB ;
\label{eq:lagr}
\eeq
where $\tilde{F}^{\mu\nu}=\epsilon^{\mu\nu\rho\sigma}F_{\rho\sigma}/2$ is the dual of the electromagnetic field ${F}^{\mu\nu}$ and
$\phi$ is the axion field. It would allow for production of axions in stars via Primakoff process~\cite{Raffelt:2006cw} and for photon-axion interconversions
in magnetic fields that make possible direct detection of axions~\cite{Irastorza:2018dyq}.

Apart from the original theoretical motivation, a renewed interest arose towards axions in the recent years since these still
elusive particles may play
a crucial role in explaining the dark matter  puzzle in the Universe~\cite{Sikivie:2006ni}.  In particular, axions with masses  in the range 10--1500 $\mu$eV would 
be the dominant cold dark matter component. 
In this context their main production mechanism would be the non-thermal realignment, and depending on the 
cosmological scenario there can be some contribution associated with decays of topological defects, like cosmic strings and domain
 walls~\cite{Ringwald:2016yge}.
Generic low-mass axion-like particles (ALPs) exhibiting a two-photon vertex, but which do not satisfy Eq.~(\ref{eq:ma}) may also be valid dark matter
 candidates~\cite{Arias:2012az}.
Despite their small mass, axions and ALPs  can be cold dark matter since being produced non-thermally they are non-relativistic and behave as 
a \emph{classical field}, exhibiting the features  of a  \emph{ Bose-Einstein condensate}~\cite{Sikivie:2009qn,Davidson:2013aba,Davidson:2014hfa}.

The presence of a two photon vertex $a\gamma \gamma$ with effective coupling $g_{a \gamma}$ makes possible
the direct detection of dark matter axions through haloscope methods. Notably, 
the
ADMX~\cite{Du:2018uak}  experiment  consists of a strong laboratory magnet which would couple the dark matter axion field with the electromagnetic  field and create microwave radiation at a detectable level. A new variation of this scheme is the dielectric haloscope approach, where the conversion of dark matter axions to microwave photons is enabled by immersing one or more dielectric layers in a strong laboratory field. This has recently been proposed through the MADMAX~\cite{TheMADMAXWorkingGroup:2016hpc} project. 
Another recent proposal is based on tunable cryogenic plasma, which  enables resonant conversion by matching the axion 
mass to a plasma frequency~\cite{Lawson:2019brd}. 

The two-photon vertex would also lead to a \emph{spontaneous axion decay}. For the KVSZ model this rate gives~\cite{Kim:1979if,Shifman:1979if,Raffelt:2006rj}
%%%%%%%%%%%%%%%%
\begin{equation}
\Gamma_{a\to \gamma \gamma} = \frac{g_{a\gamma}^2 m_a^{3}}{64 \pi} = 1.1 \times 10^{-24} \,\ \textrm{s}^{-1} \left( \frac{m_a}{\textrm{eV}}\right)^5 \,\ ,
\end{equation}
%%%%%%%%%%%%%%%
where the first term at the right-hand-side is valid for generic axion-like-particles, while the second uses the relation between $g_{a \gamma}$ and
{$F_a$}
for QCD axions. Comparing this decay rate with the lifetime of the Universe $4.3 \times 10^{17}$~s, one would realize that
for typical masses of axion dark matter these particles would be stable on cosmological scales.
However,  the presence of  background  photon radiation $f_\gamma$ might also lead to
  \emph{stimulated  axion decay}. In rest frame of the axion condensate, assuming  an isotropic photon distribution
  $f_\gamma$  much smaller than the axion one $f_a$, it reads~\cite{Caputo:2018vmy} 
%......................
\beq
\Gamma_{\rm eff}= \Gamma_{a\to \gamma \gamma} (1+2 f_\gamma) \,\ ,
\eeq
%.......................
which can increase the rate by many orders of magnitudes. It has been recently pointed out that the photons produced
by stimulated axion decays would produce a signal detectable at radio-telescopes~\cite{Caputo:2018vmy,Caputo:2018ljp,Sigl:2019pmj,Arza:2019nta,Battye:2019aco}.
In particular, an  \emph{exponential} enhancement of the photon occupation number due to stimulated axion decays has been predicted when photons 
propagate across the axion condensate with frequency close to the axion mass $m_a$. This effect has been interpreted 
as a parametric resonance of photons, leading to an instability in the axion condensate within a narrow 
frequency band $\omega \sim m_a/2$~\cite{Espriu:2011vj,Hertzberg:2018zte,Arza:2018dcy,Braaten:2016dlp}. Particularly interesting to study this phenomenon 
are systems with high axion densities, like  axion stars and 
clumps~\cite{Tkachev:1987cd,Kephart:1994uy,Hertzberg:2018zte,Arza:2018dcy}. An intriguing association between the axion instability and evaporation
of axion stars to explain
fast radio bursts has also been proposed~\cite{Tkachev:2014dpa}.
Stimulated axion decays in superradiant clouds around primordial black holes have also been recently 
characterized~\cite{Rosa:2017ury,Ikeda:2019fvj,Boskovic:2018lkj}.
Most of the studies on the subject have focussed on the photon exponential growth under the assumption of no depletion
of the axion field. A first attempt to characterise a possible depletion of the axion field has recently been
presented in~\cite{Sawyer:2018ehf,Sawyer:2019rgg}. 
We devote our work to revisit the issue of the stimulated decays of axion condensates. Starting from the Heisenberg equations
for the relevant operators, we present the kinetic equations
for the photon-axion system and study their solution, including the dynamical evolution of the axion field. We also discuss possible applications
of the axion-photon instability.

The plan of out work is as follow. In Sec.~2 we revise the basics of  spontaneous and stimulated decay of axions. 
In  Sec.~3 we study the kinetic equations for the stimulated decays of the axion  axion condensate into photons. At first we characterize the instability of the
axion system and then we present the general equations including also the feedback on the axion field. 
In  Sec.~4 we present two applications of our finding:  the case of a homogeneous axion condensate at the recombination epoch 
and the case of a localized axion clump. Finally in  Sec.~5 we summarize our results and we conclude.
{It follows an Appendix in which we discuss the conditions  for the axion instability in the expanding Universe.}

%%%%%%%%%%%%%%%%%%%%%%%%%%%
\section{Spontaneous and stimulated axion decays}
%%%%%%%%%%%%%%%%%%%%%%%%%%

%.%%%%%%%%%%%%%%%%%
\subsection{Spontaneous decay rate}
%%%%%%%%%%%%

At first we recall the derivation of the axion  \emph{spontaneous} decay rate in vacuum $a \to \gamma \gamma$. It  can be obtained from the usual perturbation theory as
%.........................
\beq
d \Gamma = \frac{1}{2 m_a} \frac{d^3 k}{(2 \pi)^3 2 E_k} \frac{d^3 p}{(2 \pi)^3 2 E_p} 
\left| {\mathcal M}_0 (a\to \gamma \gamma) \right|^2
(2 \pi)^4 \delta^4(p_a -k-p) \,\ ,  
\eeq
%........................
where the amplitude for decay is 
%%%%%%%%%%%%%%
\beq
i {\mathcal M}_0 (a\to \gamma \gamma) = i g_{a \gamma} \epsilon_\nu^\ast \epsilon_\lambda^\ast \epsilon^{\nu \lambda \alpha \beta} p_\alpha k_\beta \,\ ,  
\eeq
%%%%%%%%%%%%%%%%
where $p$ and $q$ are the momenta of the two emitted photons and $ \epsilon$ are the polarization vectors.
Including a factor $1/2$ for the phase space of identical particles, one finds
%....................
\begin{eqnarray}
\Gamma_{a\to \gamma \gamma}  &=& \frac{1}{2 m_a} \frac{1}{ 8 \pi} \frac{1}{2} \sum_{\rm pols.} \left| {\mathcal M}_0 (a\to \gamma \gamma) \right|^2 \,\ \nonumber 
\\
&=& \frac{1}{32 \pi m_a} g_{a\gamma}^2 \times 2(p\cdot k)^2 \,\ \nonumber \\
&=& g_{a\gamma}^2 \times \frac{m_a^3}{64 \pi} \,\ ,
\label{eq:decay0}
\end{eqnarray}
%......................

%%%%%%%%%%%%%%%%%%%%
\subsection{Stimulated decay rate}
%%%%%%%%%%%%%%%%%%%%

In the presence of a photon background the axion decay rate can be enhanced by a \emph{stimulation} effect. 
This can be explicitly seen from the Boltzmann equation for the axion density~\cite{Caputo:2018vmy}
{
%........................................
\begin{eqnarray}
{\dot n}_a & = & - \int d \Pi_a d \Pi_k d \Pi_p 
(2 \pi)^4 \delta^4(p_a -k-p) \left| {\mathcal M}_0 (a\to \gamma \gamma) \right|^2
[f_a(f_\gamma +1)(f_\gamma +1) - (f_a+1)f_\gamma f_\gamma \,\ \nonumber \\
&=&-\int  d \Pi_a d \Pi_k d \Pi_p (2 \pi)^4 \delta^4(p_a -k-p) \left| {\mathcal M}_0 (a\to \gamma \gamma) \right|^2
[f_a(1+2f_\gamma)-f_\gamma^2] \,\ \nonumber  \\
&\simeq& - n_a \Gamma_{a\to \gamma \gamma} (1+2 f_\gamma) 
\end{eqnarray}
%...........................................
}
where $d \Pi$ indicates the phase space Jacobian of the axion and photons. {In the last term we 
used the definition of the axion decay rate $\Gamma_a = \int  d \Pi_k d \Pi_p (2 \pi)^4 \delta^4(p_a -k-p)
\left| {\mathcal M}_0 (a\to \gamma \gamma) \right|^2/(2m_a)$ in the rest frame of the axion condensate. Moreover,
we have neglected the inverse axion decay term, proportional to $f_\gamma^2$, since for the environment we consider
$f_a \gg f_\gamma$. Under this limit, the stimulation factor only depends on $f_\gamma (p_a-k)+f_\gamma(k)$. Assuming
equal photon distributions we get the term $2 f_\gamma$.}
Therefore, we have an effective rate~\cite{Tkachev:1987cd,Kephart:1994uy}
%..........................................
\beq
\Gamma_{\rm eff} = \Gamma_{a\to \gamma \gamma}  (1+2 f_\gamma)  \,\ ,
\label{eq:stimuldecay}
\eeq
%.....................................
where the last term is the stimulation factor due to the background photon radiation.
We recall that $f_\gamma$ is  the photon occupation number which refers to a density per cell of volume $(2 \pi)^3$ (in Planckian units)
in phase space, {while we will indicate with $n_\gamma$ and $n_a$, respectively the photon and axion number densities per unit volume in the three-dimensional
space.} From the previous equation one realizes that  the effective axion decay rate into photons
 $\Gamma_{\rm eff} $ is equal to the spontaneous decay rate $\Gamma$ given by Eq. (\ref{eq:decay0})
 only if the photon occupation $f_\gamma\ll1$, and thus in particular if the final state photon modes have not been
significantly been populated by previous decays.
Otherwise $\Gamma_{\rm eff}$ can become much larger by stimulated emission.
 
%%%%%%%%%%%%%%%%%%
\subsection{Exponential photon enhancement in the narrow instability band}
%%%%%%%%%%%%%%%%%%

Now we will consider the particular case in which the enhancement factor is produced by photons with occupation numbers
{ $N_{\bf k} = N_{-\bf k} \equiv f_\gamma$}, produced by spontaneous axion decays at rest with momentum  $|{\bf k}| \simeq m_a/2$. 
In this Section we will follow an heuristic derivation that follows the treatment presented in~\cite{Mukhanov:2005sc} in the context of inflaton decays
in cosmological preheating scenarios.

 {We start by treating the axion and photon fields as a classical fields.} 
 {The axion field is real. Moreover, we take the axion  field as a non-relativistic condensate.  Therefore, taking that most of the axion dark matter is in the zero momentum mode, it reads~\cite{Davidson:2013aba,Davidson:2014hfa}
%%%%%%%%%%%
\begin{equation}
\phi= \phi_{0}\cos(m_{a}t) \,\ .
\label{eq:axionfield}
\end{equation}
%%%%%%%%%%%%
However, for further purpose, it is useful to express the real axion field
$\phi=\phi^\ast$ as non-relativistic limit of a complex scalar field~\cite{Davidson:2013aba,Davidson:2014hfa}
%...............
\beq
\phi=\frac{1}{\sqrt{2m_{a}}}\frac{1}{\sqrt{V}}\left[b_{0}e^{-im_{a}t}+b^{\ast}_{0}e^{im_{a}t}\right] \,\ .
\label{eq:axionfield}
\eeq
%.............
For a condensate, $b \simeq \sqrt{N_a}$. Therefore, $\phi_{0}=2{\textrm Re} (b_{0})/\sqrt{2m_{a} V}= \sqrt{2 n_a/m_a}$, where $n_a$ is the axion number density.
}
From the Lagrangian of Eq.~(\ref{eq:lagr}) it has been shown that working in the {Coulomb}  gauge,  where the electromagnetic potential $A^\mu = (\varphi,\bA)$ satisfies the condition $\nabla\cdot\bA=0$, {assuming an homogeneous axion field}
one can obtain the equation of motion for the vector potential (see Ref.~\cite{Wilczek:1987mv,Masaki:2019ggg} for the general equations of 
axion electrodynamics)
%%%%%%%%%%%%%%%%%%%
\beq
 \begin{split}
 &\ddot{\bA}-\nabla^{2}\bA+g_{a\gamma}\dot{\phi}\nabla\times \bA=0 \,\ .
 \end{split}
 \label{eq:vecpoteq}
 \eeq
%%%%%%%%%%%%%%%%%%%%%%
{We will consider the axion field as homogeneous also in the presence of a localized clump, assuming that the spatial field variations 
$|\nabla \phi | \ll \partial_t \phi $ as in~\cite{Hertzberg:2018zte}. Conversely, in the equations one would find a term proportional to the 
spatial gradient of $\phi$ and to the scalar potential of the electromagnetic field (see, e.g.,~\cite{Fedderke:2019ajk,Alonso-Alvarez:2019ssa}).}
We remark that for simplicity we are assuming massless photons, considering values of the axion mass much larger than the plasma frequency.
Expanding  the vector potential into plane waves 
 \beq
 \begin{split}
 \bA(t,\bx)=\sum_{\lambda}\int \frac{d^{3}\bk}{(2\pi)^3}\sqrt{\frac{V}{{2\omega_{k}}}}\epsilon_{\lambda}(\bk)c_{\bk,\lambda}e^{i\bk\cdot\bx} \,\ ,
 \end{split}
 \label{eq:planewave}
 \eeq
%%%%%%%%%%%%%%%%%
one finds the following Mathieu equation~\cite{Espriu:2011vj,Hertzberg:2018zte}
%......................................
\begin{equation}
\ddot{c}_{\bk,\pm}+k^{2}c_{\bk,\pm} \mp g_{a\gamma}\phi_{0}m_{a}k\sin(m_{a}t)c_{\bk,\pm}=0 \,\ ,
\label{eq:mathieu}
\end{equation}
%......................................
where we worked with circular polarizations ($\lambda= \pm$) expressed in terms of the polarization vectors
%%%%%%%%%%%%%%%%%%%%%%%%%%%%%
\beq
\begin{split}
\epsilon_{\pm}(\bk)&=\frac{1}{\sqrt{2}}\left(\epsilon_{1}(\bk)\pm i\epsilon_{2}(\bk)\right) \,\ ,\\
\epsilon_{1}(\bk) &=(1,0,0) \,\ , \\
\epsilon_{2}(\bk) &=(0,1,0) \,\ . \\
\end{split}
\eeq
%%%%%%%%%%%%%%%%%%%%%%%%%%%%%%%%
It is known that the Mathieu equation has instability bands where the solution shows an instability, i.e. an exponentially growing solution.
We will formally discuss the instability properties of the Mathieu equation in the next Section. Here we present a heuristic derivation.

From Eq.~(\ref{eq:mathieu}), since $k \simeq m_a/2$, we can obtain the photon dispersion relation 
{
%.....................
\begin{eqnarray}
\omega &=& k \left[1 \mp 2 g_{a \gamma} \phi_0 \sin(m_a t) \right]^{1/2} \nonumber  \\
&\simeq& \frac{m_a}{2} \left[1 \mp  g_{a \gamma} \phi_0 \sin(m_a t) \right] \,\ .
\end{eqnarray}
%................................
From this equation one obtains that }
the photons are created within a thin shell of width
%...............................
\beq
\Delta k \simeq g_{a \gamma} \phi_0 m_a \ll m_a \,\ . 
\label{eq:widhtapp}
\eeq
%.........................
Then, the number of photons created in the momentum modes around $k=m_a/2$ are given by
%...................
\begin{eqnarray}
f_\gamma(k=m_a/2) &=&  \frac{n_\gamma}{4 \pi k^2 \Delta k/ (2 \pi)^3}  \,\ \nonumber \\
&=& \frac{8 \pi^2n_\gamma }{m_a^2 g_{a \gamma} \phi_0 m_a} =
\frac{4 \pi^2 \phi_0}{m_a^2 g_{a \gamma}}\frac{n_\gamma}{n_a} \,\ .
\end{eqnarray}
%.....................
The {number density} $n_\gamma$ exceeds unity and so stimulated emission is essential only if
%................................
\begin{equation}
n_\gamma > \frac{g_{a\gamma} m_a^2}{4 \pi^2 \phi_0} n_a. 
\label{eq:nagtrng}
\end{equation}
%................................
Therefore, from Eq.~(\ref{eq:stimuldecay})  the effective decay rate reads
%.................
\beq
\Gamma_{\rm eff} = \frac{g_{a\gamma}^2 m_a^3}{64 \pi} \left(1+ \frac{8 \pi^2 \phi_0}{m_a^2 g_{a \gamma}}\frac{n_\gamma}{n_a}\right) \,\ .
\eeq
%....................
Then, the Boltzmann equation for the photons is given by
%..................
\begin{eqnarray}
{\dot n}_\gamma &=& 2 \Gamma_{\rm eff} n_a \,\ \nonumber \\
&=&   \frac{g_{a\gamma}^2 m_a^3}{32 \pi} \left(1+ \frac{8 \pi^2 \phi_0}{m_a^2 g_{a \gamma}}\frac{n_\gamma}{n_a}\right)n_a  \,\ .
\end{eqnarray}
%.............
In the case  Eq.~(\ref{eq:nagtrng}) is satisfied,  the previous equation can be easily integrated,
%................
leading to to an exponentially growing solution
%..................
\beq
n_\gamma = \exp[\tilde\mu t] n_\gamma(0)  \,\ ,
\eeq
%................
with growth rate 
%.................
\beq
\tilde\mu = \pi \frac{g_{a\gamma} m_a \phi }{4} \,\ .
\label{eq:muapprox}
\eeq
%................
Note that in this case of exponential speed-up, the growth rate is proportional to $g_{a\gamma}$
and not to $g_{a\gamma}^2$ as in the spontaneous decay.

%%%%%%%%%%%%%%%%%%%%%%%
\section{Kinetic equations for decaying axion condensates}
%%%%%%%%%%%%%%%%%%%%%

%.%%%%%%%%%%%%%%%%%
\subsection{Axion-Photon Hamiltonian}
%%%%%%%%%%%%

In this Section we will present the kinetic equations for the decaying axion condensate.
Most of the previous works treat the photon field as a classical one, and based
their solution on the Lagrangian in Eq.~(\ref{eq:lagr}). Since we will work out the quantum Heisenberg equations for the ensemble, we will work in an Hamiltonian
formalism, inspired by the approach of~\cite{Sawyer:2018ehf,Sawyer:2019rgg}. In particular, from the Lagrangian  in Eq.~(\ref{eq:lagr})
the interaction Hamiltonian is
%%%%%%%%%%%%%%%%%%
\beq
H_{\rm int}=\frac{g_{a\gamma}}{4}\int d^{3}\bx\;\phi\tilde{F}^{\mu\nu}F_{\mu\nu} =-g_{a\gamma}\int d^{3}\bx\, \phi\bE\cdot\bB\;.
\eeq
%%%%%%%%%%%%%%%%
We will work in the radiation gauge, where
the Hamiltonian assumes the form
%%%%%%%%%%%%%%%%%%
\beq
H_{\rm int}=g_{a\gamma}\int d^{3}\bx\,\phi\dot{\bA}\cdot\nabla\times\bA\;.
\eeq
%%%%%%%%%%%%%%%%%%%%%%%%
In order to quantize the Hamiltonian we expand the  vector potential  in terms of free fields, assuming the interaction is a small perturbation
%%%%%%%%%%%%%%%%%%%%%
\begin{equation}
\begin{split}
{\bf A}(t,x)&=\frac{1}{(2\pi)^{3}}\sum_{\lambda}\int d^{3}\bk \sqrt{\frac{V}{{2\omega_{k}}}}\left(\epsilon_{\lambda}(\bk)c_{\bk,\lambda}e^{-ikx}+\epsilon^{*}_{\lambda}(\bk)c^{\dagger}_{\bk,\lambda}e^{ikx}\right)\;,
\end{split}
\end{equation}
%%%%%%%%%%%%%%%%%
{where now $c_{\bk,\lambda}$ is an operator.}
Note that $Vd^3\bk/(2\pi)^3$ is the dimensionless phase space element that becomes a sum over discrete modes in a finite volume
and the operators $c_{\bk,\lambda}$ and $c^\dagger_{\bk,\lambda}$
are dimensionless annihilation and creation operators, respectively, satisfying canonical commutation relations
$[c_{\bk,\lambda},c^\dagger_{\bk^\prime,\lambda^\prime}]=(2\pi)^3\delta^{3}(\bk-\bk^\prime)\delta_{\lambda,\lambda^\prime}/V$
which reduce to $\delta_{{\bf k},{\bf k}^\prime}\delta_{\lambda,\lambda^\prime}$ in the discrete case.
Treating the axion field as a classical field for now, the quantized interaction Hamiltonian in terms of circular polarizations is then
%%%%%%%%%%%%%%%%%%
\beq
\begin{split}
H_{\rm int}=-ig_{a\gamma}\phi_{0}\cos(m_{a}t)\int_{\Omega}\frac{Vd^{3}\bk}{(2\pi)^3}\,\omega_{k}\left[\left(c_{\bk,\omega,+}c_{-\bk,\omega,+}-c_{\bk,\omega,-}c_{-\bk,\omega,-}\right)-\left(c^{\dagger}_{\bk,\omega,+}c^{\dagger}_{-\bk,\omega, +}-c^{\dagger}_{\bk,\omega,-}c^{\dagger}_{-\bk,\omega,-}\right)\right] \,\ ,
\label{eq:hint}
\end{split}
\eeq
%%%%%%%%%%%%%%%%%%%%%%%
where $c_{\bk,\omega}= c_{\bk} e^{-i \omega_{k}t}$ with $\omega_k=|k|$.
Here the integral over the phase-space only runs over half of the directions.
Finally, the total Hamiltonian can be written as
%%%%%%%%%%%%%%%%%%%%
\begin{eqnarray}
H &=&  H_{0} + H_{\rm int} \nonumber \\
&=& \sum_{\lambda= \pm}\int_{\Omega}\frac{Vd^{3}\bk}{(2\pi)^3}\,\omega_{k}\left(N_{\bk,\lambda}+N_{-\bk,\lambda}\right)-ig_{a\gamma}\phi_{0}\cos(m_{a}t)\int_{\Omega}\frac{Vd^{3}\bk}{(2\pi)^3}\,\omega_{k}\left[\left(C_{\bk,\omega,+}-C_{\bk,\omega,-}\right)-\left(C^{\dagger}_{\bk,\omega,+}-C^{\dagger}_{\bk,\omega,-}\right)\right] \nonumber \\
&=& \int_{\Omega}\frac{Vd^{3}\bk}{(2\pi)^3}\,\omega_{k}N_{\bk}-ig_{a\gamma}\phi_{0}\cos(m_{a}t)\int_{\Omega}\frac{Vd^{3}\bk}{(2\pi)^3}\,\omega_{k}\left[C_{\bk,\omega}-C_{\bk,\omega}^{\dagger}\right]
\label{eq:Hamiltot}
\end{eqnarray}
%%%%%%%%%%%%%%%%%%%%%
where we introduced the number 
%%%%%%%%%%%%%%%
\beq
N_{\pm\bk,\lambda}=c^{\dagger}_{\pm\bk,\lambda}c_{\pm\bk,\lambda} \,\ ,
\eeq
%%%%%%%%%%%%%%%
and pair correlation operators
%%%%%%%%%%%%%%
\beq
C_{\bk, \omega, \lambda}=c_{\bk, \omega, \lambda}c_{-\bk, \omega, \lambda} \,\ ,
\eeq
%%%%%%%%%%%%%%%%%%%
for a given polarization state, as well as 
the total number operator summed over polarization state
\beq
N_{\bk}=N_{\bk,+}+N_{-\bk,+}+N_{\bk,-}+N_{-\bk,-} \,\ ,
\eeq
and the difference of the pair correlation operators for the two polarization states
\beq
C_{\bk,  \omega}=C_{\bk, \omega, +}-C_{\bk, \omega, -} \,\ .
\eeq
%%%%%%%%%%%%%%%%%%%%%%%%
Note that the number operators $N_{\pm\bk,\lambda}$ do not depend on the photon frequency.
The previous operators satisfy the following commutation relations
%%%%%%%%%%%
\begin{eqnarray}
[C_{\bk,  \omega},C^{\dagger}_{\bk^\prime, \omega^\prime}] &=& (2+N_{\bk})(2\pi)^3\delta^{3}(\bk-\bk^\prime)/V \,\ , \nonumber \\
\left[N_{\bk},C_{\bk^\prime,  \omega^\prime} \right] &=& -2 C_{\bk,  \omega}(2\pi)^3\delta^{3}(\bk-\bk^\prime)/V \,\ .
\end{eqnarray}

%%%%%%%%%%%%%
\subsection{Heisenberg equations}
%%%%%%%%%%%%

From the Hamiltonian in Eq.~(\ref{eq:Hamiltot}) one obtains the following Heisenberg equations
%%%%%%%%%%%%%%%%%%%%%
\beq
i\dot{c}_{\bk, \omega, \lambda}=[c_{\bk, \omega,\lambda},H_{0}+H_{\rm int}] \,\ ,
\eeq
%%%%%%%%%%%%%%%%%%%%%%
which reads
%%%%%%%%%%%%%%%
\beq
\begin{split}
\dot{c}_{\bk,\omega,-}&=-i\omega_{k}c_{\bk,\omega,-}-g_{a\gamma}\omega_{k}\phi_{0}\cos(m_{a}t) c_{-\bk,\omega,-}^{\dagger} \,\ , \\
\dot{c}_{\bk,\omega,+}&=-i\omega_{k}c_{\bk,\omega,+}+g_{a\gamma}\omega_{k}\phi_{0}\cos(m_{a}t)  c_{-\bk,\omega,+}^{\dagger} \,\ .\\
\end{split}
\eeq
These equations can be cast under the form  of a  second order differential equation
\beq
\ddot{c}_{\bk,\omega,\pm}+\omega_{k}^{2} c_{\bk, \omega,\pm} \pm g_{a\gamma}\omega_{k}\phi_{0}m_{a}\sin(m_{a}t)c^{\dagger}_{-\bk,\omega,\pm}=0 \,\ .
\eeq
Using $c^{\dagger}_{-\bk,\omega,\pm} = -c_{\bk,\omega,\pm}$ one can rewrite the previous equations as  
%%%%%%%%%%%%%%%%%%
\beq
\ddot{c}_{\bk,\omega,\pm}+\omega_{k}^{2} c_{\bk,\omega,\pm} \mp g_{a\gamma}\omega_{k}\phi_{0}m_{a}\sin(m_{a}t)c_{\bk,\omega,\pm}=0 \,\ ,
\label{eq:mathieu1}
\eeq
%%%%%%%%%%%%%%%%%
which corresponds to the Mathieu equation found before [Eq.~(\ref{eq:mathieu})].
Here we notice that among the different applications, the Mathieu equations describes an inverted pendulum, driven 
by an external vertical pivot, oscillating at high frequency (the so called Kapitza's pendulum)~\cite{pendulum}.
The linearized equation of motion for the angular displacement $\varphi$ is given by
%.................
\begin{equation}
\ddot{\varphi} + f(1- \mu \cos t) \varphi=0 \,\ ,
\end{equation}
%...................
where $\mu$ is the driving amplitude, and $f$ is the driving frequency squared.
For such a system, a parametric
resonance can cause an instability of the pendulum, when its frequency is a multiple of the driving frequency of the pivot. 
In our system, the oscillating axion condensate acts as the external pivot driving the instability.

We will find here the exponentially growing solution of the Mathieu equation in the first instability band.
In the limit of small amplitude of the photon field, the solution of Eq.~(\ref{eq:mathieu1})
can be found analytically through a linearized stability analysis. We closely follow the discussion presented in~\cite{Hertzberg:2018zte}.
The operator $c_{\bk,\pm}$ can be expanded as
\beq
c_{\bk,\omega,\pm}=\sum_\omega e^{-i \omega t}f_{\omega,\pm}(t) \,\ ,
\eeq
where $f_{\omega,\pm}(t)$ is a slowly-varying function of $t$ and $\omega$ runs over integer multiples of $m_a/2$.
%%%%%%%%%%%%%%%
Inserting this expansion into the Mathieu equation (\ref{eq:mathieu1}) we obtain
\beq
2i\omega \dot{f}_{\omega,\pm}+(\omega^{2}-k^{2})f_{\omega,\pm}\pm\frac{g_{a\gamma}\phi_{0}m_{a}k}{2i}(f_{\omega-m_{a},\pm}-f_{\omega+m_{a},\pm})=0 \,\ .
\eeq
%%%%%%%%%%%%
Neglecting ${\ddot f_\omega}$, the lowest frequency modes corresponds
to $\omega=\pm m_a/2$ and give
%%%%%%%%
\beq
\begin{split}
&\dot{f}_{m_{a}/2,\pm}+\frac{1}{im_{a}}\left(\frac{m_{a}^{2}}{4}-k^{2}\right)f_{m_{a}/2,\pm}\mp\frac{g_{a\gamma}\phi_{0}k}{2}f_{-m_{a}/2,\pm}=0\\
&\dot{f}_{-m_{a}/2,\pm}-\frac{1}{im_{a}}\left(\frac{m_{a}^{2}}{4}-k^{2}\right)f_{-m_{a}/2,\pm}\mp\frac{g_{a\gamma}\phi_{0}k}{2}f_{m_{a}/2,\pm}=0 \,\ .
\end{split}
\eeq
%%%%%%%%%%%%%%%%%%%%
An instability is associated with an exponentially growing solution $f \sim \exp[\mu t]$. The parameter $\mu$ is called
\emph{Floquet exponent}.
One explicitly finds
%%%%%%%%%%%%%%
\beq
\mu=\sqrt{\frac{g_{a\gamma}^{2}\phi_{0}^{2}k^{2}}{4}-\frac{1}{m_{a}^{2}}\left(k^{2}-\frac{m_{a}^{2}}{4}\right)^{2}} \,\ ,
\eeq
%%%%%%%%%%%%%%%%%
where the maximum growth  obtained for $k=m_{a}/2$ is
%%%%%%%%%%%%%%%%
\beq
\mu =g_{a\gamma}m_{a}\phi_{0}/4 \,\ .
\label{eq:mumath}
\eeq
%%%%%%%%%%%

The edges of the instability band are given when $\mu=0$:
\beq
k_{L/R}=\sqrt{\frac{g_{a\gamma}^{2}\phi_{0}^{2}m_{a}^{2}}{16}+\frac{m_{a}^{2}}{4}}\pm\frac{g_{a\gamma}\phi_{0}m_{a}}{4} \,\ ,
\eeq
and the bandwidth is
%%%%%%%%%%%%%
\beq
\Delta k=g_{a\gamma}m_{a}\phi_{0}/2 \,\ .
\label{eq:band}
\eeq
%%%%%%%%%%%%%%%%%%%%

%%%%%%%%%%
\begin{figure}[t!]
\vspace{0.cm}
\includegraphics[width=0.5\textwidth]{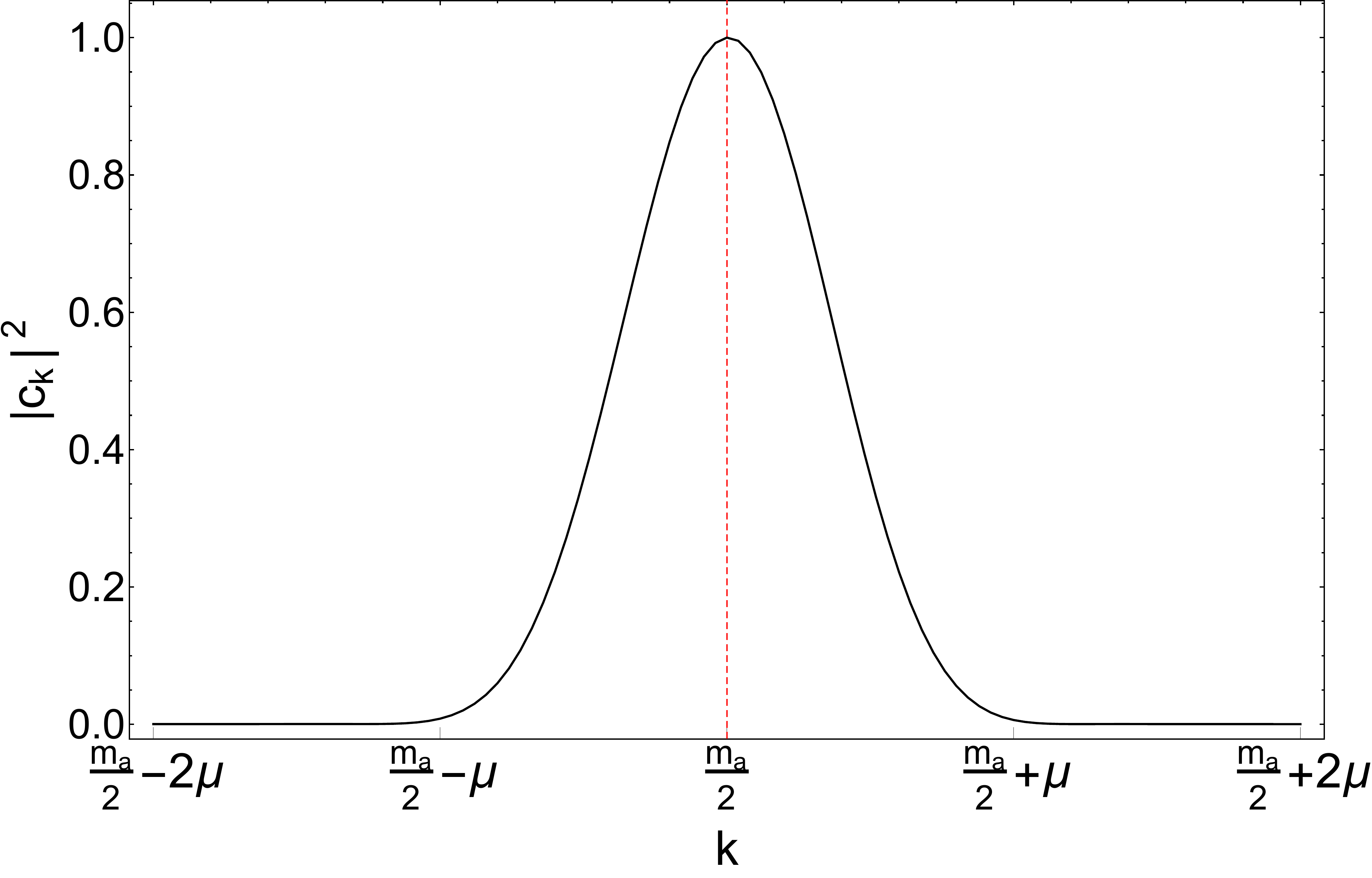}
\caption{Instability band of the Mathieu equation [Eq.~(\ref{eq:mathieu1})]. See the text for details.}
\label{fig:band}
\end{figure}
%%%%%%%%%%%%

In Figure~\ref{fig:band} we show the instability band of  the Mathieu equation [Eq.~(\ref{eq:mathieu1})] obtaining numerically  solving this equation 
and plotting at late time the behavior of $|c_{\bf k}|^2$ in function of $k$.

This solution of the Mathieu equation can be interpreted as a parametric resonance, occurring when $\omega= m_a/2$. 
This growth rate can be compared with the one 
found in Eq.~(\ref{eq:muapprox}), where we recall that $n_k \sim |c_k|^{2}$. One realizes that the two rates agree apart from 
a numerical factor. This discrepancy is due to the fact that in Eq.~(\ref{eq:widhtapp}) the approximation for the instability
width overestimated it, producing a larger photon growth rate.
From this comparison it results that the stimulated decay discussed in the previous Section can be interpreted as narrow parametric resonance
of the Mathieu equation, and vice-versa.

{Finally, we mention that the treatment of the stimulated axion decays in photon, based on the presence of an instability 
of the Mathieu equation in a narrow band, has been widely discussed in cosmology in the context of reheating after the inflation, associated 
with the decay of inflatons $\phi$ into two bosons $\chi$, i.e. $\phi \to \chi \chi$. Indeed, most of the results presented in classical papers
on the subjects (see, e.g.,~\cite{Traschen:1990sw,Shtanov:1994ce,Kofman:1997yn,Adshead:2015pva,Adshead:2018doq}), can be directly translated to the axion case.}

%%%%%%%%%%%
\subsection{Equations for Photon Numbers and Correlations}
%%%%%%%%

The Mathieu equation allows one to determine the presence of an instability band. However, it does not tell how the instability is triggered.
Indeed, one needs to assume a non-zero initial value of the $c_{\bf k}$ operator in order to have an exponential growth. In the literature,
often one introduces the concept of vacuum fluctuations to give an initial seed to $c_{\bf k}$. However, as pointed out in~\cite{Sawyer:2018ehf,Sawyer:2019rgg}, this is not
completely satisfactory, especially in a cosmological context where one should invoke a fluctuation coherent on a cosmological scale.
Moreover, the operators  $c_{\bf k}$ are not observable. Therefore,  it would be more convenient to express the equations of motion 
in terms of the photon number ${N}_{\bk}$ and correlation operators ${C}_{\bk}$. In order to obtain these sets of equations, we 
reinsert the definition of the axion field $\phi$ in terms of creation and annihilation operators $b^{\dagger}$ and $b$ [Eq.~(\ref{eq:axionfield})]
in the interaction Hamiltonian $H_{\rm int}$ of Eq.~(\ref{eq:hint}). Moreover, since we will consider only modes in the instability band
with $\omega_k \approx m_a/2$, removing the rapidly oscillating terms in the Hamiltonian, we get
%%%
\beq
%H_{\rm int} = -i \frac{\mu}{\sqrt{N_a}}\left(\frac{m_a}{2}\right)^3\int_{|{\bf k}|= \frac{m_a}{2}} Vd^2{\hat{\bf n}}\,\left[b^{\dagger}C_{{\bf k}}-bC^{\dagger}_{{\bf k}}\right] \,\ ,
H_{\rm int} = -i \frac{\mu}{\sqrt{N_a}}\sum_{|{\bf k}|= \frac{m_a}{2}}\left[b^{\dagger}C_{\bk}-bC^{\dagger}_{\bk}\right] \,\ ,
\eeq
%%%%%%%
where the sum is extended on modes  with  $|{\bf k}|= m_a/2$.
From the previous Hamiltonian, working in a \emph{co-rotating reference system} with frequency $m_a/2$ we find the following Heisenberg equations
%%%%%%%%%%%%%%%%
\beq
\begin{split}
\dot{N}_{\bk}&=2\kappa\left[b^{\dagger}C_{\bk}+bC^{\dagger}_{\bk}\right] \,\ ,\\
\dot{C}_{\bk}&= \kappa(2+N_{\bk})b \,\ ,\\
\dot{b}&= -\kappa\sum_{|{\bf k}|= \frac{m_a}{2}}C_{\bk} \,\ ,
\end{split}
\label{eq:quantumeq}
\eeq
%%%%%%%%%%%%%%
which agree with the kinetic equation recently presented in~\cite{Sawyer:2018ehf,Sawyer:2019rgg} and where the
combination $\kappa\equiv\mu/\sqrt{N_a}$ is manifestly independent of $N_a$, as it should be for an operator equation
involving $b$.

In the following we will consider the solution of the previous system in the \emph{mean field} approximation, assuming that the previous operators
are $c$-numbers. A  discussion about effects beyond mean field has been recently given in~\cite{Sawyer:2018ehf,Sawyer:2019rgg}.

Note that the Hamiltonian of Eq.~(\ref{eq:quantumeq}) has a nice analogy with the \emph{trilinear Hamiltonian} used in quantum optics
to explain \emph{spontaneous parametric down-conversion}~\cite{Mollow:1967zza,Bonifacio:1970zz,bandilla}. Indeed if we limit ourselves to the case of a single mode with $\omega_k=m_a/2$ 
we get
%%%%%%%%%%%%%%
\beq
H_{\rm int}=\kappa \left[b^{\dagger}c_{\bk}c_{-\bk}+b c^{\dagger}_{\bk}c^{\dagger}_{-\bk}\right] \,\ ,
\label{eq:parametricdown}
\eeq
%%%%%%%%%%%%%
where $c_{\bk,+}c_{-\bk,+}-c_{\bk,-}c_{-\bk,-} \equiv c_{\bk}c_{-\bk}$.
In the language of quantum optics this Hamiltonian describes the interaction of the pump $b$
with two photons: the  signal $c_{\bk}$ and the idler $c_{-\bk}$.

%%%%%%%%%%%%%%%%%%%%%%%%%%%
\subsubsection{Short-time evolution}
%%%%%%%%%%%%%%%%%%

At first, we consider the short time solution,   assuming that the axion field has not been depleted by the 
decays into photons. In this limit, assuming a constant axion field with $b \approx b^{\dagger} \approx \sqrt{N_{a}}$
the Heisenberg equations for the previous photon number operator and the correlations read
%%%%%%%%%%%%%%%%%%%%%%%%%%%%%%%
\beq
\begin{split}
\dot{N}_{\bk}&=2\mu\left[C_{\bk}+C^{\dagger}_{\bk}\right] \,\ ,\\
\dot{C}_{\bk}&= \mu(2+N_{\bk}) \,\ ,\\
\end{split}
\eeq
%%%%%%%%%%%%%%%%%%
where $\mu=g_{a\gamma}m_{a}\phi_{0}/4$ is the growth rate of Eq.~(\ref{eq:mumath}), found from the Mathieu
equation.
It is interesting that in these quantum equations, even we assume that initially there is no radiation
 photon $N_{\bf k} =0$, one can still have a non trivial evolution. Indeed, $\dot{C}_{\bk} \neq 0$,
 since in the last term in the right-hand-side of the second equation there is still an initial photon 
 component $(2+ N_{\bf k}) \simeq 2$ due to the spontaneous axion decay. Even without
 pre-existing photon background, the spontaneous axion decay provides an unavoidable photon seed.
 
One finds
%%%%%%%%%%
\beq
\ddot{N}_{\bk}=4\mu^{2}(2+N_{\bk}) \,\ ,
\eeq
%%%%%%%%%
whose solution is 
%%%%%%%%%%%
\beq
 N_{\bk}(t)=N_\bk(0)e^{2\mu t}+2(\cosh(2\mu t)-1) \,\ .
 \label{eq:nkshort}
\eeq
%%%%%%%%%%%

%%%%%%%%%%
\begin{figure}[t!]
\vspace{0.cm}
\includegraphics[width=0.5\textwidth]{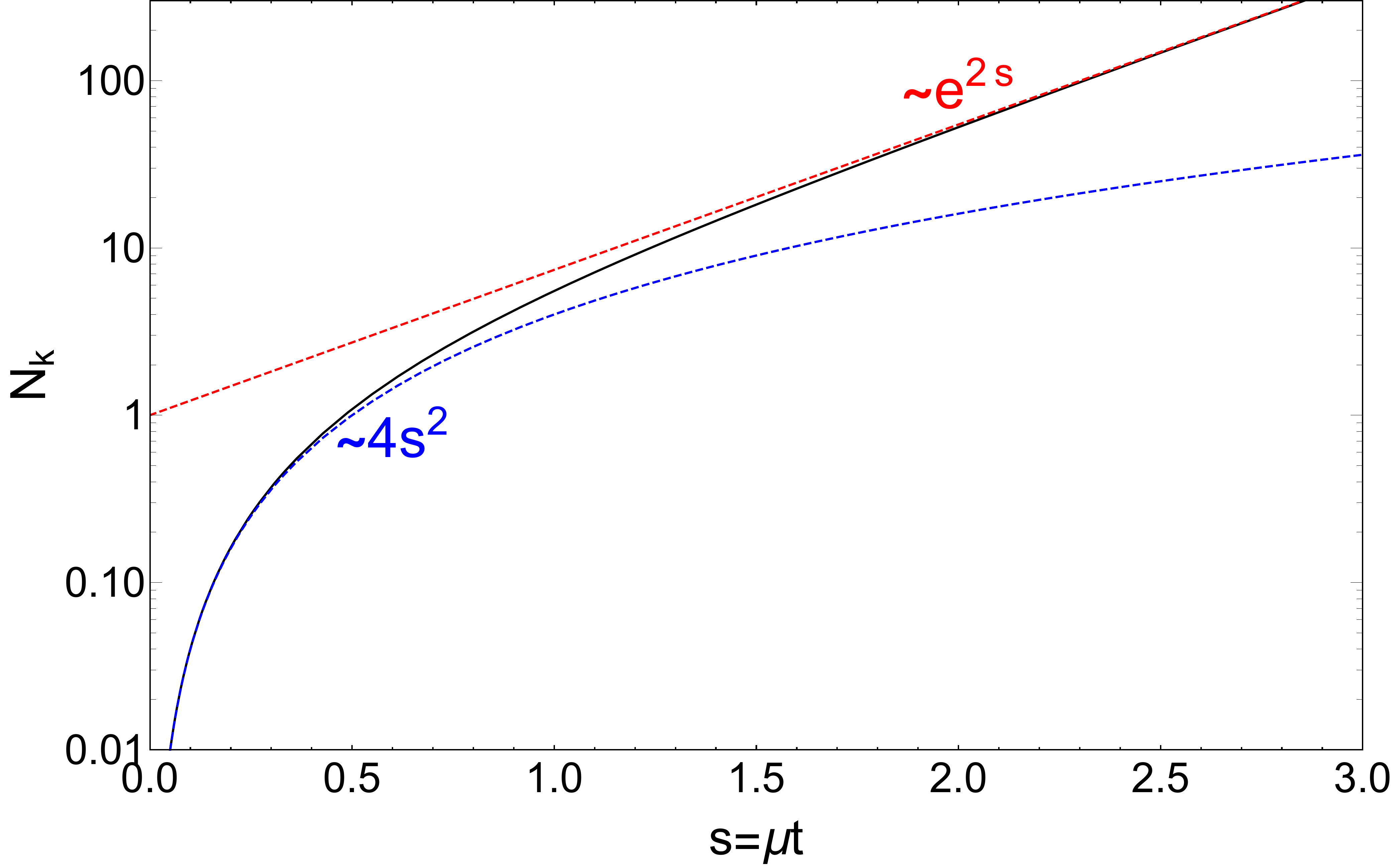}
\caption{Short-time solution of axion-photon conversions of Eq.~(\ref{eq:nkshort}) for $N_\bk(0)=0$.}
\label{fig:limit}
\end{figure}
%%%%%%%%%%%%

In Figure~\ref{fig:limit} we show the behavior of  the previous equation as a function of $s=\mu t$, assuming $N_\bk(0)=0$.
It is interesting to notice that for $t \ll \mu^{-1}$ the previous equation has as solution
$ N_{\bk} \sim 4 s^2 $. This corresponds to the initial phase where the spontaneous decay 
scaling as $g_{a \gamma}^2$ populates the photon field. Instead the long-time solution  for
$t \gg \mu^{-1}$ scales as $N_{\bk} \sim \exp[2 s]$, in agreement with the exponential solution
found from the Mathieu equation.
In the language of quantum optics this limit corresponds to a parametric amplifier in the limit of a constant pump.
{In the cosmological environment, which is the natural case for the effect we are studying, there are many sources 
of photons with energy $\omega=m_a/2$ which can trigger the stimulated axion decays, e.g. a fraction of the photons of
the cosmic microwave background or a fraction of the photons in normal matter at some temperature will have
some photons of energy $\omega=m_a/2$. Therefore, the presence of an initial photon seed  $N_{\bk}\neq 0$ is guaranteed. 
Nevertheless, it is intriguing to realize that even assuming  $N_{\bk}=0$, the decay of a single axion on the entire universe would be
enough to trigger the stimulated photon emission, due to the high occupation number of the axion condensate. 
For comparison, the stimulated decays of inflatons after inflations, discussed in preheating scenarios, 
needs vacuum fluctuations to provide the initial seed for the further decays~\cite{Traschen:1990sw,Shtanov:1994ce}.
 }

%%%%%%%%%%%%%
\subsubsection{Back-reaction on axions}
%%%%%%%%%%%%%

%%%%%%%%%%
\begin{figure}[t!]
\vspace{0.cm}
\includegraphics[width=0.5\textwidth]{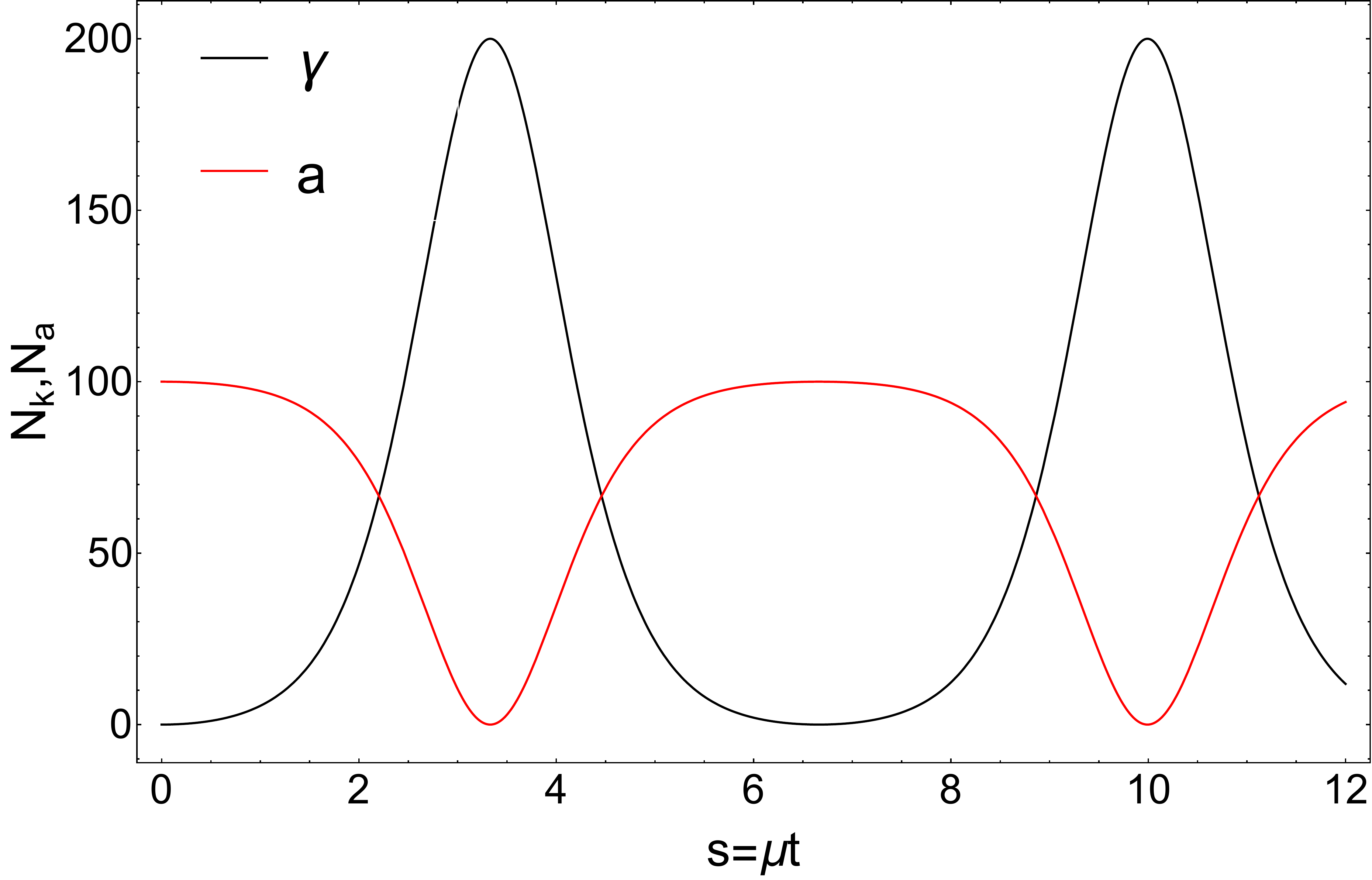}
\caption{Evolution of $N_a$ and $N_{\bf k}$ in function of $s=\mu t$ for the mode with $\omega_k=m_a/2$ for an initial
$N_a=100$. }
\label{fig:evol}
\end{figure}
%%%%%%%%%%%%

We consider now the full solution of the kinetic equations of motion
[Eq.~(\ref{eq:quantumeq})] taking into account also the evolution of the axion field.
We consider the case of $\omega_k=m_a/2$. In Fig.~\ref{fig:evol} shows the evolution of the 
axion occupation number for an initial $N_a=100$, as a function of the variable
$s = \mu t$. We find that the axion condensate exhibits a periodic behavior, passing through minima of total depletion 
and maximum of repopulation (anticorrelated with the photon population) with a period $T=2/\mu$.
In order to develop an analytical interpretation of the results discussed before, we follow the treamtment presented in~\cite{Bonifacio:1970zz}
for a trilinear Hamiltonian, exhibiting spontaneous parametric down conversion. We consider the simplified version
of the Hamiltonian, described by Eq.~(\ref{eq:parametricdown}), adding also the kinetic term for photons and axions
%%%%%%%%%%%%%
\begin{eqnarray}
H_{\rm int}&=& \kappa \left[b^{\dagger}c_{\bk}c_{-\bk}+b c^{\dagger}_{\bk}c^{\dagger}_{-\bk}\right] \,\ , \nonumber \\
H_{0}&= & m_{a}b^{\dagger}b+\frac{m_{a}}{2}\left(c^{\dagger}_{\bk}c_{\bk}+c^{\dagger}_{-\bk}c_{-\bk}\right) \,\ .
\end{eqnarray}
%%%%%%%%%%%%%%%%%
We 
 introduce the operators
\beq
R^{+}=b^{\dagger}c_{-\bk}\,\ , \quad |R|^{2}=R(R+1) \,\ ,  \quad R_{3}=\frac{1}{2}\left(b^{\dagger}b-c^{\dagger}_{-\bk}c_{-\bk}\right) \,\ ,  \quad R=
\frac{1}{2}\left(b^{\dagger}b+c^{\dagger}_{-\bk}c_{-\bk}\right) \,\ .
\eeq
which satisfy the algebra of the angular momentum operators.
Using these latter the Hamiltonian reads
\begin{eqnarray}
H_{\rm int}&=\kappa\left[R^{+}c_{\bk}+R^{-}c^{\dagger}_{\bk}\right] \,\ , \nonumber \\
H_{0}&=m_{a}(R_{3}+c^{\dagger}_{\bk}c_{\bk}) \,\ ,
\end{eqnarray}
after  subtracting a constant term
\beq
C=m_{a}\left(\frac{1}{2}b^{\dagger}b+c^{\dagger}_{-\bk}c_{-\bk}-\frac{1}{2}c^{\dagger}_{\bk}c_{\bk}\right)
\eeq
from the free Hamiltonian.
The equations of motion have the form
\beq
\dot{c}_{\bk}=-i \kappa R^{-} \,\ , \quad \dot{R}^{-}=2i  \kappa c_{\bk}R_{3}  \,\ , \quad  \dot{R}_{3}=i  \kappa (c_{\bk}^{\dagger}R^{-}-c_{\bk}R^{+})  \,\ .
\eeq
We now make the mean field approximation which should be justified for reasonably large photon occupation numbers and we obtain
%%%%%%%%%%
\beq
\dot{\gamma}_{\bk}=  \kappa r_{1}  \,\ , \quad \dot{r}_1=2  \kappa \gamma_{\bk}r_{3}  \,\ , \quad \dot{r}_{3}=-2  \kappa \gamma_{\bk}r_{1} \,\ ,
\eeq
%%%%%%%%%%
where $-i\gamma_{\bk}=\langle c_{\bk}\rangle$, $r_{1}=\langle R^{\pm}\rangle$.
The constants of motion are
%%%%%%%%%%%%
\beq
r_{3}^{2}+r_{1}^{2}=J^{2}  \,\ ; \quad M=\gamma^{2}_{\bk}+r_{3} \,\ , 
\eeq
and 
%%%%%%%%%%%%%%%%%
\beq
\dot{\gamma}_{\bk}^{2}=  \kappa^{2} r_{1}^{2}=J^{2}-(M-\gamma_{\bk}^{2})^{2} \,\ .
\eeq
Introducing the polar coordinates
\beq
r_{1}=J\sin\theta  \,\ , \quad r_{3}=J\cos\theta \,\ ,
\eeq
one gets 
\beq
\dot{\theta}=2\kappa \gamma_{\bk}\rightarrow \ddot{\theta}=2  \kappa^{2}  J\sin\theta \,\ .
\label{eq:pendulum}
\eeq
The parameters J and $\theta$ can be read as
%%%%%%%
\begin{eqnarray}
J&=& \frac{1}{2}(N_{a}+N_{-\bk})=r \,\ , \\
\theta &=& \arctg\left(\frac{r_{1}}{r_{3}}\right)=\arctg\left(\frac{2\sqrt{N_{a}N_{-\bk}}}{N_{a}-N_{-\bk}}\right) \,\ .\label{eq:theta}
\end{eqnarray}
%%%%%%%%%%%%%%%%%%
Eq.~(\ref{eq:pendulum}) represents the evolution of an 
\emph{inverted pendulum} with frequency $\sqrt{2 J}$. 
The period of the pendulum is given by~\cite{landau}
%.................................
\beq
T= \frac{4}{\sqrt{2J}}K(\sin(\theta_{0}/2)) \,\ ,
\eeq
%..................................
where the function $K$ is the complete elliptical integral
%......................
\beq
 K(k)=\int_{0}^{\pi/2}d\xi\frac{1}{\sqrt{1-k^{2}\sin^{2}\xi}} \,\ ,
\eeq
%........................
and for an inverted pendulum $\theta_0 \to \tilde{\theta} = \theta_0+ \pi$, with $\theta_0 \ll 1$
since initially $N_{-\bk} \simeq 0$. 
Expanding the elliptical integral we obtain the period of the pendulum as
%.................
\beq
T\simeq\frac{{2}}{\mu}\left|\ln\left(\frac{N_{-\bk}}{N_{a}+N_{-\bk}}\right)\right|\simeq\frac{{2}}{\mu}\ln N_{a} \,\ ,
\label{eq:period}
\eeq
since $\cos^{2}(\theta_{0}/2)=\frac{1}{2}-\frac{1}{2}(N_{a}-N_{-\bk})/(N_{a}+N_{-\bk})$. 
Therefore we reproduce the result seen in Fig.~\ref{fig:evol} and we found a logarithmic correction associated with 
$N_a$ as predicted in~\cite{Sawyer:2018ehf}. 
It is interesting to notice that the behavior of inverted pendulum of the stimulated emission of photons has a nice
analogy with the behavior of a dense neutrino gas under flavor conversions~\cite{Hannestad:2006nj}.

Next we consider the fact that axions may decay in different directions. 
It is easy to realize that increasing the number of directions one reduces the logarithmic delay in the period~\cite{Sawyer:2018ehf}
%%%%%%%%%%%%%%%%%%
\beq
T=\frac{{2}}{\mu}\left[\ln\left( \frac{N_{a}}{N_{d}}\right)\right] \,\ .
\eeq
%%%%%%%%%%%%%%%%%%%%%%
Indeed,  considering our system enclosed in a  box of size $L$, so that the wave-numbers are quantized 
as
%%%%%%%%%%%%%%%%%%%
\beq
\bk=\frac{2\pi}{L}{\bf n}\;;
\eeq
%%%%%%%%%%%%%%%%%%
where ${\bf n}$ is a vector of integers. The number of possible directions, given the constraint $\omega_{k}=m_{a}/2$, is
\beq
 \quad N_{d}=\int d^{3}{\bf n}\,\delta\left(n-\frac{Lm_{a}}{4\pi}\right)=4\pi \int dn\,n^{2}\delta\left(n-\frac{Lm_{a}}{4\pi}\right)=4\pi n^{2}\bigr|_{\omega_{k}=m_{a}/2}=\frac{1}{4\pi} L^{2}m_{a}^{2}\;,\label{eq:N_d}
\eeq
where $n=|{\bf n}|$.
Finally we relax the assumption that $\omega_{k}=m_{a}/2$ and we consider possible unstable modes inside the band
of Eq.~(\ref{eq:band}).  
The number of modes in this band is
%%%%%%%%%%%%%%%%%%
\[
\frac{2\pi}{L}\Delta n=2\mu\rightarrow \Delta n=\frac{\mu L}{\pi}
\]
%%%%%%%%%%%%%%%%%%%%%%%%%%%
Then, the effective axion number in Eq.~(\ref{eq:period}) is reduced  by another factor $N_{t}=\mu L/\pi$.
Therefore, including also this effect we find 
%................
\beq
T=\frac{{2}}{\mu}\left[\ln\left( \frac{N_{a}}{N_{d} N_t}\right)\right] \,\ .
\label{eq:logdel}
\eeq
%...........

%%%%%%%%%%%%%
\section{Cosmological applications}
%%%%%%%%%%%%%

We analyse here the possibility of the axion-photon instability in two cosmological environments: the Early Universe at recombination and 
a localized high-density axion clump.

%%%%%%%%%%%%%
\subsection{Early Universe}
%%%%%%%%%%%%%

At first we consider an homogeneous axion field at recombination 
(at redshift $z=1100$). The density of axion cold dark matter
 is given by~\cite{Sikivie:2009qn}
 %%%%%%%%%%%%%%%%%%
\beq
\begin{split}
n_{a}(t_{0})&=\left(\frac{4\times10^{47}}{\cm^{3}}\right)\left(\frac{F_{a}}{10^{12}\GeV}\right)^{5/3}\left(\frac{a(t_{1})}{a(t_{0})}\right)^{3}=\\
&=\left(\frac{1.1\times10^{14}}{\cm^{3}}\right)\left(\frac{g_{a\gamma}}{10^{-11}\GeV^{-1}}\right)^{-5/3} \,\ ,
\end{split}
\eeq
%%%%%%%%%%%%%%%%%%%%%%%
where where $a(t_{1})$ corresponds to the scale factor at the QCD phase transition ($T_{1}=1\GeV$), when axions are produced,
and we expressed the axion-photon coupling as~\cite{Arias:2012az}
%%%%%%%%
\beq
g_{a\gamma}=\frac{1}{2\pi}\frac{\alpha}{F_{a}} \,\ ,
\eeq
%%%%%%%%%%
where $\alpha$ is the fine-structure constant.
In the numerical evalution we used a photon-axion coupling possible for generic ALP dark matter.
For QCD axions, one should consider a coupling smaller by three order of magnitues.
The typical exponential growth rate of photons from stimulated axion decays
is $\mu \sim g_{a\gamma}\sqrt{m_{a}n_{a}}$. In order to have an efficient production, at first
one should require that $\mu > H$, where $H$ is the Hubble parameter. 
In  the matter-dominated era  $H=H_{0}(1+z)^{3/2} $ with  $H_{0}=68 \km\s^{-1}\Mpc^{-1} =2.2\times 10^{-18}\textrm{s}^{-1} $
The photon plasma frequency can be neglected after recombination because $\omega_{p}< 10^{-11}\eV\ll m_{a}$. One finds
%%%%%%%%%%%%%%%%%%%%%%%
\beq
\begin{split}
&\mu /H \simeq \left(\frac{4\times10^{47}}{\cm^{3}}\right)^{1/2}\left(\frac{F_{a}}{10^{12}\GeV}\right)^{-1/6}\left(\frac{a(t_{1})}{a(t_{0})}\right)^{3/2}\\
&\frac{\alpha}{2\pi}10^{-12}\GeV^{-1}\left(\frac{m_{a}}{\mu\eV}\right)^{1/2}\mu\eV^{1/2}\frac{1}{H_{0}(1+z)^{3/2}}=\\
&=\left(\frac{F_{a}}{10^{12}\GeV}\right)^{-1/6}\left(\frac{m_{a}}{\mu\eV}\right)^{1/2}\frac{1.4\times10^{9}}{(1+z)^{3/2}}=\\
&=2.0\times10^{10}\left(\frac{g_{a\gamma}}{10^{-11}\GeV^{-1}}\right)^{1/6}\left(\frac{m_{a}}{10^{-5}\eV}\right)^{1/2}(1+z)^{-3/2} \,\ .
\end{split}
\eeq
%%%%%%%%%%%%%%%%%%%%%%%%%%
This first condition is satisfied. However, as noticed already in seminal papers  \cite{Preskill:1982cy,Abbott:1982af}
the expansion of the Universe would redshift momenta of previously created photons  taking them out the instability 
layer. Therefore, the occupation numbers relevant for the stimulated conversions are smaller than in a static case.
The number of photons in the instability band is given by $\Delta k/ m_a \sim \mu /m_{a}$. Therefore in order 
to have an efficient photon growth one should require~\cite{Preskill:1982cy,Abbott:1982af}
%%%%%%%%%%%%%%%%
\beq
\frac{\mu}{m_a} \frac{\mu}{H} \gg 1 \,\ .
\label{eq:cosmocond}
\eeq
%%%%%%%%%%%%%%%%%%
However, in our case we numerically obtain
%%%%%%%%%%%%%%%%%%%%%%
\beq
\begin{split}
\frac{\mu}{H}\frac{\mu}{m_{a}}&=2.1\times10^{-9}\left(\frac{F_{a}}{10^{12}\GeV}\right)^{-1/3}(1+z)^{-3/2}=\\
&=4.2\times10^{-8}\left(\frac{g_{a\gamma}}{10^{-11}\GeV^{-1}}\right)^{1/3}(1+z)^{-3/2} \,\ ,\label{eq:cosmo}
\end{split}
\eeq
%%%%%%%%%%%%%%%%%%%%%%
Therefore, based on this argument, already seminal papers~\cite{Preskill:1982cy,Abbott:1982af}
excluded a sizable decay of the axion dark matter (see, however, the recent discussion in~\cite{Sawyer:2019rgg}). 
We notice that a similar  analysis has been presented in the context of inflaton reheating during 
reheating~\cite{Traschen:1990sw,Shtanov:1994ce,Kofman:1997yn}. 
{A formal derivation of the Mathieu equation characterizing the axion instability [Eq.~(\ref{eq:mathieu})] in the expanding Universe and
the effect of the redshift is presented in Appendix A.}

%%%%%%%%%%%%%
\subsection{Axion Clumps}
%%%%%%%%%%%%%

If the Peccei-Quinn symmetry is broken after inflation, the axion field may present large inhomogeneities from one Hubble patch to another one.
In this case gravitational interaction would organize axion dark matter in the form of  localized \emph{clumps}, known as Bose stars, oscillons, that
can arrange themselves in miniclusters~\cite{Kolb:1993zz}. These dense axion objects have motivated investigations on the stimulated decay of axions.
As an example, we consider the clump  model presented in~\cite{Schiappacasse:2017ham} which has the following parameters 
%%%%%%%%%%%%%%%%%%%%%%%%%%%
\beq
\begin{split}
N_{\rm max}&=1.6\times10^{56}\left(\frac{m_{a}}{10^{-5}\eV}\right)^{-2}\left(\frac{g_{a\gamma}}{10^{-11}\GeV^{-1}}\right)^{-1} \,\ , \\
R_{\rm min}&=6.5\times10^{5}\km\left(\frac{m_{a}}{10^{-5}\eV}\right)^{-1}\left(\frac{g_{a\gamma}}{10^{-11}\GeV^{-1}}\right) \,\ ,\label{eq:clumps}\\
\rho_{\rm max}&=2.4\times10^{-18}\kg\cm^{-3}\left(\frac{m_{a}}{10^{-5}\eV}\right)^{2}\left(\frac{g_{a\gamma}}{10^{-11}\GeV^{-1}}\right)^{-4} \,\ ,
\end{split}
\eeq
%%%%%%%%%%%%%%%%%%%
where $N_{\rm max}$ is the maximum axion number, $R_{\rm min}$ is the minimum radius, and $\rho_{\rm max}$ is the maximum axion density 
to guarantee stable configurations. 
For these parameter one obtains as photon exponential growth rate 
%%%%%%%%%%%%%%%%%
\beq
\mu=\frac{1}{\sqrt{8}} g_{a\gamma}\rho_{\rm max}^{1/2}=1.7 \times 10^{-9} \textrm{km}^{-1}\left(\frac{m_{a}}{10^{-5}\eV}\right)\left(\frac{g_{a\gamma}}{10^{-11}\GeV^{-1}}\right)^{-1}\label{eq:clumps2}
\eeq
%%%%%%%%%%%%%%%%%%%
{Note that for an axion clump the effect of the redshift due to Universe expansion is negligible in comparison to the 
gravitational field of the clump itself.}
This growth  rate $\mu$ should be  compared with a typical size $R$ of the clump~\cite{Hertzberg:2018zte}. 
For an homogeneous axion clump one should require $\mu  R  \gg 1$ in  
order to achieve 
a significant stimulated photon emission with an axion feedback. In the inhomogeneous case, as long as the growth rate is adiabatic
%......................
\beq
\left|\frac{1}{\mu} \frac{d \mu}{dr}\right| \ll \mu \,\ ,
\eeq
%..........................
 one would get as e-folding factor for the photon amplitude $ \int \mu(r) dr$ (see Eqs.~(B31)-(B32) in~\cite{Shtanov:1994ce}). However, from the adiabaticity condition since $d \mu / dr \sim \mu/R$ one gets once more
$\mu  R  \gg 1$ for a sizable effect, such that our rough estimate is justified. 
{Therefore, for our benchmark axion parameters, in order to get $\mu  R  \gg 1$ one should have an axion minicluster of size $R \gtrsim 10^3 R_{\rm min}$. In this situation the minicluster might even  ``explode'' under
photon emission~\cite{Tkachev:2014dpa}.
We also mention that in~\cite{Arza:2018dcy} it has been estimated that axion minicluster might reach an extension $R$
for which $\mu R\gtrsim1$ for $g_{a\gamma}\gtrsim2\times10^{-13}\,[m_a/(10^{-5}\,{\rm eV})]^{1/3}\,{\rm GeV}^{-1}$.}

\subsection{Constraints}

We note that in general globally only a very small fraction $\Delta N_a/N_a$ of axion dark matter can be converted into photons because
the energy in radiation today is much small than the rest mass density in dark matter, $\Omega_{\rm dm}$ in units of the critical density~\cite{Sigl:2019pmj}.
More quantitatively, if axions form a fraction $f_{\rm dm}$ of the cold dark matter, then
\beq
  \frac{\Delta N_a}{N_a}f_{\rm dm}\lesssim\frac{\Omega_\gamma(m_a/2)}{\Omega_{\rm dm}}\frac{\Delta\nu}{\nu}\,,\label{eq:constraint}
\eeq
where $\Omega_\gamma(m_a/2)$ is the photon energy density per logarithmic photon energy in terms of the
critical density and $\Delta\nu/\nu$ is the maximum of the predicted axion conversion line width and experimental relative
frequency resolution, provided that no line is detected. 
For frequencies $\nu\lesssim1\,$GHz the energy density in the universal radio
background is of the order of $2\times10^{-8}\,{\rm eV}{\rm cm}^{-3}$, see
e.g. Ref.~\cite{Hill:2018trh}, which should be compared with the dark matter density
$\Omega_{\rm dm}\rho_{\rm crit}\simeq10^3\,{\rm eV}{\rm cm}^{-3}$. Therefore,
for $m_a\sim\mu$eV, $\Omega_\gamma(m_a/2)/\Omega_{\rm dm}\sim10^{-10}$.

The number of converted axions will be $\Delta N_a\sim N_dN_t[N_{m_a/2}(T)-N_{m_a/2}(0)]$, where $N_k(t)$ is given by
Eq.~(\ref{eq:nkshort}), $N_dN_t$ is the number of photon modes into which the axion can decay, see Eq.~(\ref{eq:N_d})--(\ref{eq:logdel}),
and $T$ is the timescale over which the resonance acts which is typically of the order the size $R$
of the axion system. Since $N_a$ is a huge number, a significant constraint
is only obtained for $\mu T\gg1$. The above constraint thus simplifies to
\beq
  \mu T\lesssim\frac{1}{2}\ln\left[\frac{\Omega_\gamma(m_a/2)}{f_{\rm dm}\Omega_{\rm dm}}
  \frac{N_a}{N_dN_t(N_{m_a/2}(0)+1)}\frac{\Delta\nu}{\nu}\right]\,,\label{eq:constraint2}
\eeq
This is generally fulfilled for the cosmological case and the axion clumps discussed above.
It is, however, clear
that Eq.~(\ref{eq:constraint2}) excludes that a significant fraction of axions is in the form of structures satisfying
$\mu T\gg1$, independent of any external photon field. This could constrain axion miniclusters~\cite{Arza:2018dcy}.

Eq.~(\ref{eq:constraint2}) is different from constraints discussed \cite{Sigl:2019pmj} which considered the relative enhancement
of a diffuse photon background impinging on an axion condensate in the Mathieu regime.

%%%%%%%%%%%%%%%
\section{Conclusions}
%%%%%%%%%%%

In this work we revisited the stimulated decay of 
dark matter axion condensates  into photon pairs. In particular, we have considered the decay into a narrow instability width around photon energies
$\omega \approx m_a/2$ where one can have
 an exponential growth of the photon occupation number.
 This effect has been often interpreted as a parametric resonance of photons from the axion-photon coupling,
treating the photon field as a classical one and limiting the solution to the initial exponential growth. 
 We have analysed in details this effect presenting a mean field solution of the axion-photon kinetic equations, in terms of number of photons
 and pair correlations. We have studied the limit of negligible axion depletion, recovering the known instability. Moreover, we extended the results including
 a possible depletion of the axion field.
 Interestingly, in the mean field approximation we find that the solution can be interpreted in terms of the periodic motion of an inverted pendulum. 
 
The instability of axion condensates to decay into photons is relevant for cosmological environments. In particular, we considered the case of
a homogeneous axion field at recombination and a localized axion clump. In the first case, we confirm that due to the expansion 
of the Universe the photon production would not be efficient, since modes will quickly be redshifted away from the instability band. 
Conversely, in a localized clump one might have relevant photon production. We also discussed constraints that could result from overproduction
of the diffuse photon background by axion-photon conversion which exclude that any significant fraction of dark matter axions is in structures with $\mu T\gg1$.
This can concern the case of axion miniclusters of relatively large size for which the feedback can be significant.
Further studies are  necessary to investigate this  case. Other systems to which our formalism can be applied include the case 
of evaporation of axion stars that can lead to fast radio bursts~\cite{Tkachev:2014dpa}, and
stimulated axion decay in superradiant clouds around primordial black holes~\cite{Rosa:2017ury,Ikeda:2019fvj,Boskovic:2018lkj}.
The study of these interesting systems is left to a future work.

\section*{{Note added}}
{After our preprint appeared, in~\cite{Alonso-Alvarez:2019ssa} a similar study has been presented. The two works reach similar conclusions.}

\section*{Acknowledgements}
{A.M. thanks Raymond Sawyer and Javier Redondo for stimulating discussions.
We thank Georg Raffelt for useful comments on the manuscript.}
We acknowledge Cannon Vogel for discussions of the Kapitza pendulum.
The work of P.C. and 
A.M. is partially supported by the Italian Istituto Nazionale di Fisica Nucleare (INFN) through the ``Theoretical Astroparticle Physics'' project
and by the research grant number 2017W4HA7S
``NAT-NET: Neutrino and Astroparticle Theory Network'' under the program
PRIN 2017 funded by the Italian Ministero dell'Universit\`a e della
Ricerca (MUR). The work of G.S. was supported under Germany's Excellence Strategy - EXC 2121 "Quantum Universe" - 39083306.
We thank Cannon Vogel for discussions of the Kapitza pendulum.

%%%%%%%%%%%%%%%%%%%%%%
\section*{Appendix A: Axion instability in the expanding Universe}
%%%%%%%%%%%%%%%%%%

In this Appendix we generalize the  treatment to the  axion-photon instability in the case of the expanding Universe, characterized
by the Friedman-Robertson-Walker (FRW) metric, whose line element reads
\begin{equation}
ds^{2} = dt^2-a(t)^2g_{ij}dx^i dx^j \,\ ,
\end{equation}
where $a(t)$ is the scale factor of the Universe, $g_{ij}$  is the three metric for a flat three-space.

The ALP-photon Lagrangian is
\beq
\mathcal{L}=-\frac{1}{4}g^{\mu\alpha}g^{\nu\beta}F_{\mu\nu}F_{\alpha\beta}-\frac{g_{a\gamma}}{4}\phi \frac{1}{2}\frac{\epsilon^{\mu\nu\alpha\beta}}{\sqrt{-g}}F_{\mu\nu}F_{\alpha\beta} \,\ ,
\eeq
and the equations of motion for the electromagnetic field are given by
\beq
\partial_{\mu}\frac{\partial(\sqrt{-g}\mathcal{L})}{\partial(\partial_{\mu}A_{\nu})}=\frac{\partial(\sqrt{-g}\mathcal{L})}{\partial A_{\nu}} \,\ .
\eeq 
Working in the Coulomb gauge ($\partial_i A^i=0$) and assuming an homogeneous axion field, one gets the modified Maxwell equations 
%.............
\beq
\left(\frac{\partial^2}{\partial t^2} +  H \frac{\partial}{\partial t}- \frac{\nabla^2}{a^2}
\right)\bA -
 g_{a\gamma}{\dot \phi}(t)\frac{\nabla}{a}\times\bA=0 \,\ , 
 \label{eq:wavered}
\eeq
%..................
where we  introduced the Hubble parameter
$H={\dot a}/{a}$.~\footnote{Note that in terms of  conformal time $d\eta=dt/a$ and
  comoving coordinates the wave operator in Eq.~(\ref{eq:wavered}) can be written as $(\frac{\partial^2}{\partial \eta^2}-\nabla^2)$.}

Expanding the electromagnetic field in plane wave as in Eq.~(\ref{eq:planewave}) and considering that the axion field of Eq.~(\ref{eq:axionfield}) scales with Universe expansion 
as $a^{-3/2}$ we get the following equation of motion for the two polarization states of photons (see also~\cite{Alonso-Alvarez:2019ssa}):
%...........................
\beq
\frac{d^2 c_{\bf k, \pm}}{d x^2} + \frac{2 H}{m_a}\frac{d c_{\bf k, \pm}}{dx} + \left(A_{\bf k} \mp 2 q \cos x \right) c_{\bf k, \pm}=0 \,\ ,
\eeq
%.....................
where {$x= m_a t/2-\pi/2$}, $A_{\bf k}= 4 |\bk|^{2}/ m_{a}^{2}a^2$ and $q=2g_{a\gamma}\phi_{0}|\bk|/m_{a} a^{5/2}$.
The previous equation is consistent with Eq.~(A1) of Ref.~\cite{Alonso-Alvarez:2019ssa}.
In the case $H \ll m_a$, which is satisfied in the situations we consider, the friction term can be neglected and one obtains a Mathieu equation, as in our 
Eq.~(\ref{eq:mathieu}). A discussion
of the solution of this equation is given in~\cite{Kofman:1997yn,mathieueq} to which we address the interested readers for further details. 
An important feature of the solution of this equation is the presence of an exponential instability
$ c_{\bf k, \pm} \sim \exp (\tilde\mu x)$. In the case of a narrow resonance $q \ll 1$, the instability occurs in narrow
bands around $A_{\bf k}= n^2$, with $n=1,2, \ldots$. Each band in momentum space has width
$\Delta k \sim q^n$, so for $q < 1$  the widest and most important instability
band is the first one, $A_{\bf k} \sim 1\pm q$, where  the maximum growth rate is  ${\tilde \mu}= q/2$.
In our specific case we find 
as resonance frequency $k/a=m_{a}/2$, and growth rate 
\beq
\mu= \tilde\mu  \frac{m_a}{2} =\frac{g_{a\gamma}\phi_{0} k}{2 a^{5/2}} \,\ .
\eeq
The edges of the instability band are given by
\beq
\frac{\Delta k}{a}=2 \mu \,\ .
\eeq

As discussed in~\cite{Kofman:1997yn} in the context of cosmological preheating  an important mechanism which can prevent
the axion instability from being efficient is the redshift
of momenta $k$ away from the instability band. 
The total width of the instability band is $\sim \frac{1}{2} q m_a$.
The time $\Delta t$ during 
which a given mode remains within this band  can be estimated as $q H^{-1}$.
During this  time the number of photons
in growing modes increases as $\exp\left(\frac{q^2 m}{2 H} \right)$. This leads to an efficient growth of photons 
only if 
%..................
\beq
q^2 m \gtrsim H \,\ ,
\eeq
%...................
that corresponds to the condition we gave in Eq.~(\ref{eq:cosmocond}).

\end{document}